\documentclass[sigconf]{acmart}
\usepackage{fancyhdr}
\usepackage[utf8]{inputenc} 
\usepackage[T1]{fontenc}    
\usepackage{nicefrac}       
\usepackage{microtype}      
\usepackage{helvet}  
\usepackage{courier}  
\usepackage{url}  
\usepackage{graphicx}  
\usepackage{color}
\usepackage{bm}
\usepackage{enumitem}
\usepackage{amsmath}
\usepackage{multirow}
\usepackage{subcaption}
\usepackage{booktabs}
\usepackage{array}
\usepackage{diagbox}
\usepackage{booktabs}
\usepackage{soul}
\usepackage[ruled]{algorithm2e}
\usepackage{algorithmic}
\usepackage{textcomp}

\newcommand{\vpara}[1]{\vspace{0.01in}\noindent\textbf{#1 }}
\newcommand{\figref}[1]{Fig.~\ref{#1}}
\newcommand{\eqnref}[1]{Eq.~(\ref{#1})}
\newcommand{\defref}[1]{Definition.~\ref{#1}}
\newcommand{\secref}[1]{Sec.~\ref{#1}}
\newcommand{\tableref}[1]{Table~\ref{#1}} 

\newtheorem{definition}{Definition}

\newcommand{\ratios}{\textit{conversion index}}

\newcommand{\tcode}{\textit{Taocode}}
\newcommand{\tcodes}{\textit{Taocodes}}
\newcommand{\taobao}{\textit{Taobao}}
\newcommand{\modelname}{\textit{InfNet}}
\newcommand{\numofbaseline}{8}
\newcommand{\modelN}{\textit{Structural block}}
\newcommand{\modelE}{\textit{Diffusion block}}
\usepackage{balance} 
\AtBeginDocument{%
	\providecommand\BibTeX{{%
			\normalfont B\kern-0.5em{\scshape i\kern-0.25em b}\kern-0.8em\TeX}}}

\copyrightyear{2021}
\acmYear{2021}
\setcopyright{acmcopyright}\acmConference[SIGIR '21]{Proceedings of the 44th
International ACM SIGIR Conference on Research and Development in Information
Retrieval}{July 11--15, 2021}{Virtual Event, Canada}
\acmBooktitle{Proceedings of the 44th International ACM SIGIR Conference on
Research and Development in Information Retrieval (SIGIR '21), July 11--15, 2021,
Virtual Event, Canada}
\acmPrice{15.00}
\acmDOI{10.1145/3404835.3462898}
\acmISBN{978-1-4503-8037-9/21/07}

\begin{document}
\fancyhead{}

\title{How Powerful are Interest Diffusion on Purchasing Prediction: A Case Study of Taocode}


\author{Xuanwen Huang$^{\dagger}$, Yang Yang$^{*\dagger}$, Ziqiang Cheng$^\dagger$, Shen Fan$^{\S}$, Zhongyao Wang$^{\S}$\\ Juren Li$^\dagger$, Jun Zhang$^{\S}$, Jingmin Chen$^{\S}$}
\affiliation{%
\institution{$^{\dagger}$Zhejiang University, $^{\S}$Alibaba Group}
}
\affiliation{%
\institution{\{xwhuang,  yangya, petecheng, jrlee\}@zju.edu.cn\\ \{fanshen.fs, zhongyao.wangzy, zj157077, jingmin.cjm\}@alibaba-inc.com}
}
\thanks{*Corresponding author}
%
%

\renewcommand{\shortauthors}{Huang, et al.}

\ccsdesc[500]{Information systems~Online shopping}
\ccsdesc[500]{Information systems~Electronic commerce}
\ccsdesc[300]{Information systems~Social networks}
\begin{abstract}
	A \tcode~ is a kind of specially coded text-link on \taobao.com (the world's biggest online shopping website), 
	through which users can share messages about products with each other.
    Analyzing \tcodes~can potentially facilitate understanding of the social relationships between users and,
	more excitingly, their online purchasing behaviors under the influence of \tcode~ diffusion.
    This paper innovatively investigates the problem of online purchasing predictions from an information diffusion perspective, with \tcode~ as a case study. 
	Specifically, we conduct profound observational studies on a large-scale real-world dataset from \mbox{\taobao}, containing over 100M \mbox{\tcode~}sharing records.
	Inspired by our observations,
	we propose \modelname, a dynamic GNN-based framework that models the information diffusion across \tcode. 
	We then apply \modelname~ to 
	item purchasing predictions.
	Extensive experiments on real-world datasets validate the effectiveness of \modelname~ compared with \numofbaseline~ state-of-the-art baselines. 
\end{abstract}

%

\keywords{purchasing prediction, information diffusion, GNN, \tcode}



\maketitle

\section{Introduction}

Online shopping has become an increasingly common practice for thousands of households.
As reported by \textit{Alibaba Group}\footnote{See the FY 2020 annual report on \url{https://www.alibabagroup.com/en/ir/reports}.},
in the fiscal year of 2020,
the mobile MAUs across the China retail marketplaces comes to 846 million, 
and the total GMV of \mbox{\taobao}\footnote{the biggest online shopping website in the world that belongs to Alibaba Group.} reaches 3,387 billions of CNY.
Due to the extremely large scale of online retail marketplaces,
it has become crucial for e-commerce platforms to better understand users' online purchasing behaviors.
One classical but challenging problem in this area is that of purchasing predictions~\cite{grbovic2015commerce}:
given a user, along with a specific product or a category of goods, the question is how likely the user would be to buy this item.

Existing works have made great efforts in this field. 
Traditional methods mainly consider users and products by matrix factorization, 
statistical learning models~\cite{sismeiro2004modeling}, etc.
Deep learning-based models adopt deep neural networks to integrate the diverse information in the e-commerce system~\cite{xu2019relation,zhou2019deep,wang2020beyond,wu2020diffnet++}, 
such as users' clicking, browsing and purchasing histories. 
Most of these previous frameworks utilize the historical information of both users and products; 
however, few of them have studied the impact of information diffusion~\cite{yang2015rain,yang2016social} across the social network on users' online purchasing behaviors~\cite{ma2011recommender,ye2012exploring,ma2009learning}. 
Taking a concrete example in practice, 
a user will be more likely to buy a product at a time when she actually needs it, 
or alternatively, when some close friends recommend that item to her 
(even though she might not have any urgent demand for it).
Furthermore, such preferences for different products can spread across  online social networks: 
a user who receives a recommendation from her friends may share this item with other users within her social circle~\cite{song2006personalized}, thus promoting further possible purchases, 
while the scale of sharing could explode as the message is propagated throughout the social relationships between users. 
We refer to this phenomena as diffusion of user interests in products,  i.e., ``\textit{interest diffusion}'' for short,  
which is likely to influence the purchasing behavior of users 
to some degree.
%

Although such social factors are expected to be highly relevant to users' purchasing decisions,
they have been rarely studied in previous works for two main reasons.
First, interest diffusion across online e-commerce platforms is always \textit{implicit} and \textit{hard to observe}. 
While we can easily obtain users' browsing, purchasing and product review histories, 
or even discussions on public social media platforms, 
the majority of recommendations and interest diffusion occur between acquaintances via private social communications,
which are difficult to discover and analyze.
Indeed, previous works have explored the impact of product sharing on users' shopping behavior through analysis of email data~\cite{portrait,viral}; 
however, the social cost of using emails to share goods is much higher than using instant social tools, 
meaning that these studies may be biased in real-world scenarios. 
Second, interest diffusion across social networks is very complicated to model, 
since diffusion trajectories change naturally at any time with the spread of social interactions,
and on online shopping platforms, user interest in different products could exhibit various diffusion patterns. 
Modeling the dynamics of interest diffusion remains a challenging research problem.
   
\begin{figure}[t!]
	\centering
	\includegraphics[width=0.45\textwidth]{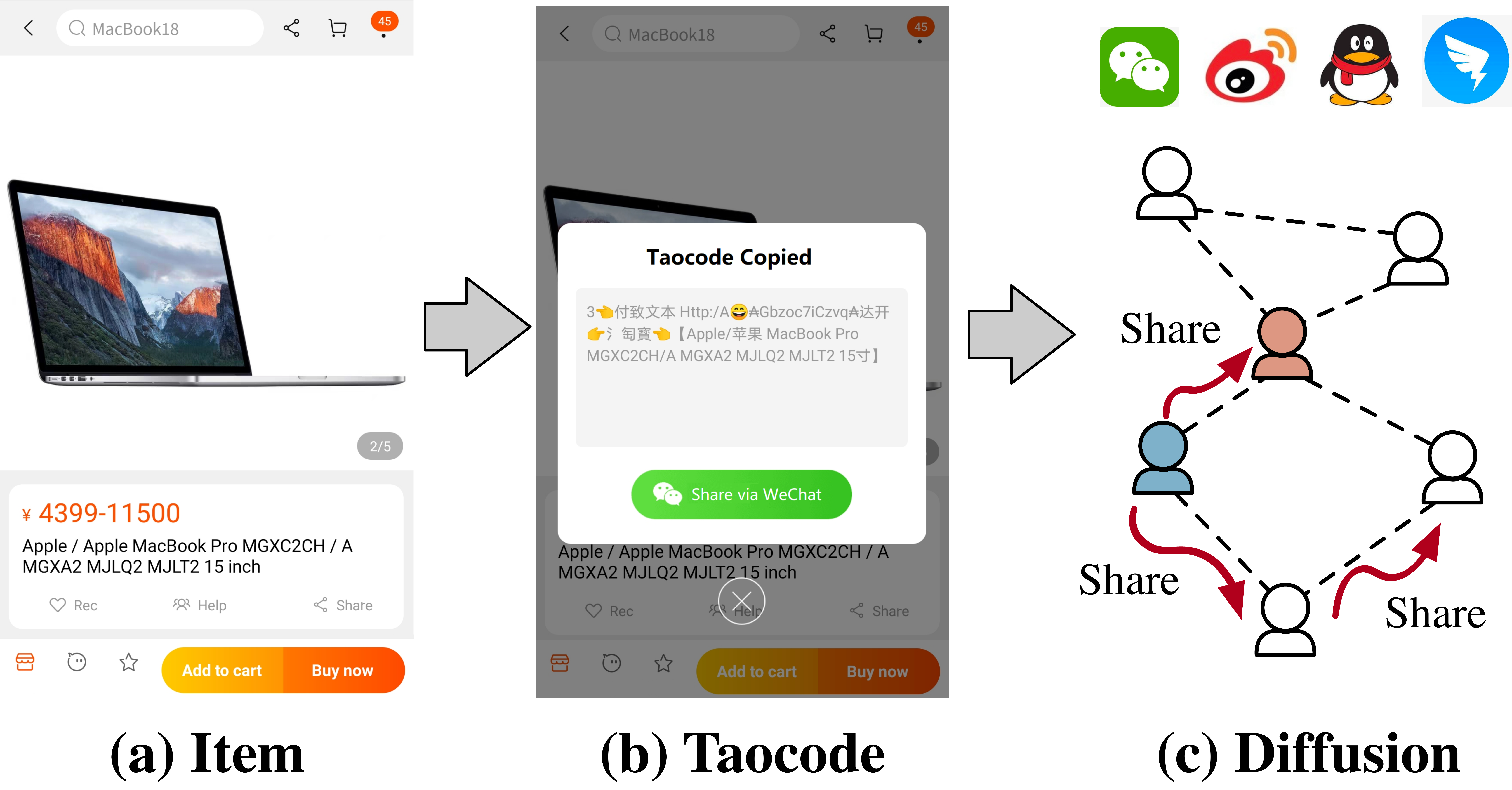}
	\caption{An illustration of \tcode~diffusion process.
		\small
		When users browse an item on~\taobao~(a),
		they can share its homepage to friends by generating the \tcode~message (b),
		and sending it via social media.
		The \tcode~ then spread over the social network (c).
		\normalsize
	}
	\label{fig:tcode case}
\end{figure}


Excitingly, \tcode, a kind of specially coded text-link on \taobao,
provides us with an appropriate testbed to observe and capture the social influences and product interest diffusion between users in the online shopping scenario.
\tcode~ is essentially a hyper-link containing the information of some specific product on \taobao.
\figref{fig:tcode case} illustrates how the \tcode~is created and spread.
It is convenient for users to create and share a \tcode~ by clicking specific options in \taobao~ when they browse an item, 
and anyone who receives the \tcode~ can browse this product by opening the embedded hyper-link.
More importantly, since \tcodes~ spread within \taobao,
the diffusion process of each sharing message can be recorded in its entirety;
we can trace the \textit{explicit sharing path} of the item on the diffusion flow, 
and obtain the details of which user shares which product with whom at what time.
This provides us with a great opportunity to model the dynamics of product sharing diffusion.
%

This paper presents a large-scale study on the effect of product interest diffusion on online shopping platforms,
taking \tcodes~sharing on \taobao~as an example.
It is pioneering work from the perspective of user interest diffusion to predict item-level purchases.
We sample a total of 100M \mbox{\tcode}~ sharing records, 
and then construct the \textit{interest diffusion networks} from those \tcode~records at different time steps:
there is a unique diffusion network within each single time span,
where each node represents a user, 
and the directed edges between two nodes denote the relationships of \tcode~ sharing. 
Moreover, the dynamics of the network structures reveal the message diffusion across the edges,
which may reflect the purchasing preferences of users regarding different groups of items.
Given a specific query of a user-item pair,
our goal is to estimate how likely the user is to buy this product based on purchasing and \tcode~ diffusion histories,
by utilizing the sequence of interest diffusion networks during the given time span. 
%

As discussed above, modeling interest diffusion networks remains challenging:
it is difficult for deep sequential models to handle geometric data,
while most existing GNNs may be unable to deal with the dynamic edge weights at different times.
To deeply understand how \tcode~messages are spread over the social network, 
we first conduct empirical observations across the entire \tcode~ diffusion (\secref{sec:observe}), 
which provides several comprehensive insights:
1) \tcodes~sharing behaviors are indeed highly correlated with users' purchasing decisions,
in that users who send or receive more \tcodes~
are likely to make more purchases;
2) the influence of \tcode~ diffusion varies across different product categories, e.g., 
items with higher price index are more likely to be purchased after being shared via \tcodes;
3) the graph structures of interest diffusion networks reflect users' purchasing preferences:
for instance, users whose neighbors share \tcodes~ more actively tend to have higher purchasing rates;
4) temporal factors play an important role in \tcode~ diffusion, i.e., 
the time interval between \tcode~senders buying and sharing the item has implicit impacts on the receivers. 

Motivated by these findings, along with the advances of sequence- and graph-level attentions,
we propose \modelname~ to integrate both the structural and temporal information of dynamic attributed diffusion networks. 
The key hints derived from our observations are that,
on one hand, a user's preference for different items is highly related with her own and her neighbors' interest diffusion patterns,
while on the other hand, those patterns are linked with time and are affected by neighborhoods' dynamic product interests at different time steps.
In other words, the structural and temporal dynamics of interest diffusion change in an interrelated fashion,
and should thus be incorporated into a unified framework.
Accordingly, we adopt a multi-level attention mechanism to model the interest diffusion patterns,
i.e., two graph attentions on diffusion networks,
and a self-attention-based encoder on diffusion sequences.
Extensive experiments applying purchasing predictions to real-world datasets
on \taobao~ demonstrate the effectiveness of our proposed model:
\modelname~ significantly outperforms \numofbaseline~ state-of-the-art baselines, 
ranging from session-based models to GNN-based social recommendation frameworks. 
We also conduct ablation studies to determine how different levels of attention work,
and visualize several cases to illustrate the diffusion patterns of products extracted by \modelname.
Overall, the contributions of this paper can be described as follows:
\begin{itemize}[leftmargin=*]
	\item We present a large-scale study on the effect of product sharing in online shopping scenarios,
	and investigate the problem of online purchasing predictions from an information diffusion perspective. 
	More specifically, we take \tcode, a specially coded type of text link of \taobao, as a case study.
	\item Comprehensive observations over 100M \tcode~sharing records illustrate several implicit characteristics of product interest diffusion across the social network on \taobao.
	\item Motivated by our observational insights, we design an end-to-end framework, named \modelname, to make item purchasing predictions. 
	Experimental results validate the effectiveness of the way, in which we adopt multi-level sequential and graph attentions to capture the users' preferences for different products and the interest diffusion dynamics. 
\end{itemize}

\section{Problem formulation}\label{sec:preliminaries}
To clearly illustrate the proposed framework focusing on the purchasing predictions from the perspective of information diffusion, 
we first give the detailed and formal definitions of interest diffusion based on \tcode~ sharing and the problem formulation.

We refer to the set of users and products in our dataset from \taobao~ as $\mathcal{V}$ and $\mathcal{P}$ respectively,
and further denote $t_i \in T=\{t_1, t_2, \cdots, t_n\}$ as the time step; 
here, $n$ is the number of the observed time step,
and each $t_i = (t_b^i, t_e^i)$ represents a specific period of time, starting at $t_b^i$ and ending at $t_e^i$.

\begin{definition}
	\textbf{\tcode~ diffusion.}
	Each \tcode~ sharing message represents a diffusion record, 
	which we formulate as a 4-tuple $d \in \mathcal{D}_t = \{(u, v, p, t_d)\ | u, v \in \mathcal{V}, p \in \mathcal{P}, t=(t_b, t_e) \in T\}$, 
	satisfying $t_b \le t_d \le t_e$.
	Here, $u$ and $v$ denote the receiver and sender respectively of the \tcode~message,
	$p$ is the item that is shared, and $t_d$ is the time at which the \tcode~ is received by $u$ from $v$.
	
\end{definition}

\begin{definition}
	\textbf{Interest diffusion network}
	is a directed attributed network $\mathcal{G_D} = (\mathcal{V}, \mathcal{E}, \mathcal{H} | \mathcal{D})$
	constructed from a \tcode~ diffusion set $\mathcal{D}$,
	where each node $u \in \mathcal{V}$ represents an online shopping user with feature vector $\mathcal{H}_u$,
	while the edge $e \in \mathcal{E}$ indicates the interest diffusion messages between users,
	with the features $H_e$ denoting the edge attributes.
	Formally, there exists an edge $e_{uv}$ between two users $u$ and $v$, 
	if the tuple $(u, v, p^*, t^*)$ is an element in $\mathcal{D}$ for some item $p^*$ and timestamp $t^*$,
	and the attribute $H_{e_{uv}}$ of the edge $e_{uv}$ contains the detailed information of the product $p^*$, e.g., PI, categories, etc.
	\label{def:idn}
\end{definition}

\begin{definition}
	\textbf{Dynamic interest diffusion network} is a sequence of interest diffusion networks $\mathcal{G} = \{\mathcal{G}_1, \cdots, \mathcal{G}_n\}$ 
	constructed from \tcode~diffusion sets $\mathcal{D} = \{\mathcal{D}_1, \cdots, \mathcal{D}_n\}$, where $i$ indicates the time step $t_i$.
	For each time step $t$, we collect all \tcode~ diffusions $\mathcal{D}_t$ that occurs during the given time span $(t_b, t_e)$ from the data source,
	and add edges into the network $\mathcal{G}_t$ as described in \defref{def:idn}.
\end{definition}

\textbf{Problem definition.} 
Considering a dynamic interest diffusion network $\mathcal{G}$, along with a query $q = (v,p,t_i)$ that $v\in\mathcal{V}, p\in\mathcal{P}, t_i\in T$, 
the problem we aim to solve is that of determining whether the user $v$ will purchase product $p$ within the time span $t_i$. 
Since the total number of items on an online shopping platform (\taobao.com in our experiment) is very large, 
to simplify the problem and investigate the influence of \tcode~diffusion,
we implement a restriction that $v$ should receive \tcode~messages containing the product $p$ from some other users during the time span $t_{i - 1}$.
We do not place limitations on that whether the user $v$ has received the \tcode~ before $t_{i - 1}$;
thus, both the short- and long-term impact of \tcode~diffusion are expected to be modeled to achieve better performance.
The additional benefit is that the number of possible queries is significantly reduced,
which makes it feasible to conduct comparison experiments between selected baselines.
\section{Observational studies}
\label{sec:observe}

\begin{figure*}[ht]
	\centering
	\includegraphics[width=1.0\textwidth]{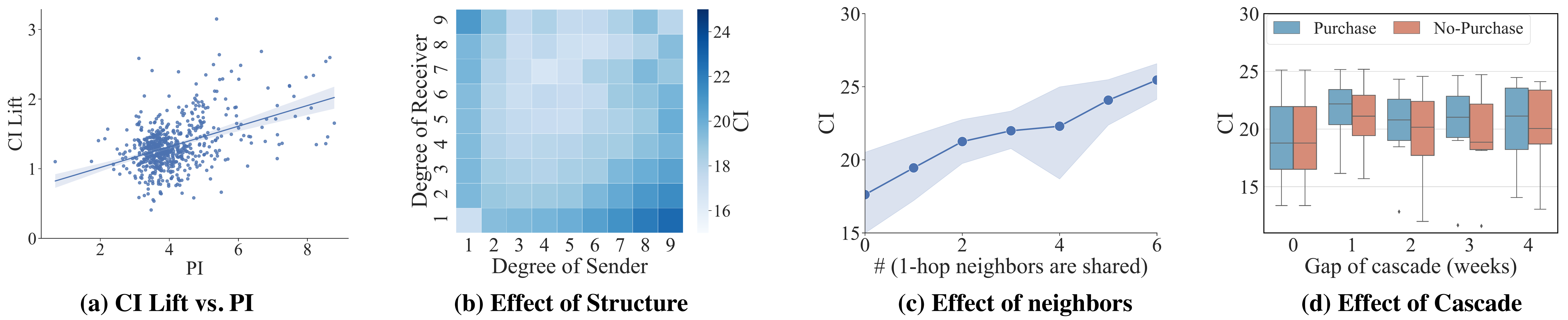}
	\caption{Observational studies on \tcode~ diffusion. 
	\small
	(a) illustrates the positive correlations between the PI and CI Lift of items, indicating that \tcode~has a stronger influence over expensive goods.
	Both (b) and (c) reflect the structural properties of \tcode~diffusion; for example, the numbers of \tcode~messages being sent and received by a user and her neighbors would affect her purchasing behavior.
	(d) finds that \tcode~messages could have a long-term impact on users' purchases.
	\normalsize	}
	\label{fig:ob:all}
\end{figure*}

In this section, we provide an in-depth examination of a large-scale dataset of \tcode~ sharing records from three different perspectives; this is done to facilitate a better understanding of \tcode~ diffusion's intrinsic properties, along with its influences on the users' purchases.

\subsection{Dataset descriptions}
The dataset for our observational studies is described as follows:
we sample records of \tcode~ sharing between \taobao~ users during a specific period of time in 2020.
The entire dataset contains over 100 million \tcode~sharing records.
Each record consists of basic information, including:
1) who creates and shares the \mbox{\tcode}, namely the \textit{sender};
2) who receives the \mbox{\tcode}, namely the \textit{receiver};
3) when the receiver opens the \mbox{\tcode~}link, namely the \textit{timestamp},
and 4) the details of the item that this \mbox{\tcode~} represents, including its \textit{Price Index} (PI)\footnote{Price Index, defined as $\mbox{PI}=f(\mbox{price})$, reflecting the price level of the product, where $f(\cdot)$ is a monotonically increasing non-linear scalar mapping due to privacy concerns.} and the category it belongs to.


To determine the correlations between sharing and purchasing, we extract the purchasing logs related to the \tcode~sharing records.
In fact, there is no explicit evidence to indicate that a user buys an item as a direct result of receiving the \tcode~;
thus, we adopt the implicit influences of \tcode~ by collecting the purchasing records of the receiver and the item that each \tcode~message contains, 
and assume that \tcode~ sharing is the main reason that leads to the purchases if the receiver buys that item \textit{after} receiving a \tcode~ within a predefined time span.
Note that for the purchasing predictions, 
we collect another group of \tcode~samples,
and the above-mentioned dataset is only used for the observations, feature extractions for users and items, along with the \tcode~diffusion network construction.
We will introduce the ``prediction dataset'' for the purchasing predictions in \secref{sec:exp}.

\subsection{\tcode's effect on purchases} 
\label{subsec:observe:effect}
The first research question is:
\textbf{Does \tcode~diffusion have impacts on users' purchasing behaviors? (Q1)}
To directly answer the question, we introduce the \ratios~(CI):
\begin{definition}

The Conversion Index (CI) of an item $p$ is the proportion of its purchasing records to the total, i.e.,
$$f(\frac{\mbox{\#(purchasing records of $p$)}}{\mbox{\#(all records of $p$)}})$$
$f$ is a monotonically increasing scalar function that maps the origin proportion to another value for data privacy issues.
It can be formulated under different conditions: (1) CI of browsing, where records are browsing histories; (2) CI of \tcode, where records are \tcode~ messages.
We further denote the ratio of \mbox{\tcode~}sharing CI and browsing CI, i.e., $\frac{\mbox{CI(\tcode)}}{\mbox{CI(sharing)}}$,
as \textit{CI Lift}, to reveal the effect of \mbox{\tcode~}sharing on purchasing behaviors.
If an item has a high CI Lift, it is much more likely that it will be purchased after being shared via \mbox{\tcode}.

\end{definition}

We compare the averaged CI of both the \tcode~ sharing and product browsing records for all products. 
The first of our findings is that items shared via \mbox{\tcode~} are much more likely to be purchased: 
a 31.0\% higher CI on average (p-value$<$0.01) than that with browsing records alone.
Meanwhile, we group the items by their leaf categories, count the averaged PI among each group, and compare their CI Lift. 
As shown in \figref{fig:ob:all}(a), there is a statistically significant positive correlations between CI Lift and averaged PI (slope=0.0448, p-value<0.01). The Spearman correlation between PI and CI Lift is 0.336  (p-value$<$0.01).
In other words, as the item's PI increases, its CI Lift becomes higher; 
i.e., \tcode~sharing can further promote its purchase rate,
which indicates that when buying expensive goods, users often engage in more consultation before making a final decision.
As the only condition we control is that of whether the products are shared via \tcode,
we conclude that \textbf{\tcode~diffusion does indeed affect users' purchasing behaviors}:
users will be more willing to make a purchase if they receive the \tcode~message from someone else.

\subsection{Structural dynamics of \tcode} 
\label{subsec:observe:structural}
We next investigate how \tcode~ affect the purchases in the following two subsections, starting with the following question: 
\textbf{Are there any structural characteristics of \tcode~diffusion patterns? (Q2)}
We first examine the structural properties of \tcode~diffusion networks.
We merge all edges among each dynamic interest diffusion network $\mathcal{G}_t$ at every time step $t$ into one graph $\mathcal{G}$,
and calculate the sum of the in and out degree with respect to different item categories for each user.
Since more than 80\% users have a degree between 1 and 9, 
and the degree distribution exhibits a long-tail shape, 
we only count \tcodes~ where the degrees of both sender and receiver lie between 1 and 9. 
The results are shown in \figref{fig:ob:all}(b):
the average CI of all \tcode~ records is 19.0, reaching a maximum of 22.7 when the sender's degree is 9 and the receiver's is 1,
and 20.9 when the degrees of the sender and receiver are 1 and 9 respectively. 
It is consistent with previous findings in the literature that ``consumers who are central in networks are quite susceptible to others' influences''~\cite{lee2010the}.
However, CI of \tcodes~ where the degrees of both sender and receiver are 1 or 9 are only 17.2 and 17.9 ( much lower than average). 
Contradicting the initial assumption that \tcode~ sharing between high degree users would have a higher CI,
we find that when the gap of degrees between the sender and receiver is larger, CI becomes higher. 
This may be explained by noting that a user's online purchases are likely to happen more frequently 
if she interacts socially with another user who has a very different pattern of social behaviors. 
For example, if a user $u$ who often shares products with others, sends a \tcode~message to user $v$ that who uses \tcode~ relatively infrequently,
one possible motivation for this sharing is that the item is of urgent demand for $v$;
thus, $v$ is more likely to make a purchase than usual.

Next, we investigate the influence of the receiver's neighbors in the \tcode~diffusion.
More specifically, for a \tcode~message $(u, v, p, t)$, we look at $v$'s 1-hop neighbors who have received and bought the same product $p$, denoted as \textit{close neighbors}. 
The intuition here is that, the number of \textit{close neighbors} of the receiver can reflect the popularity of the item during the diffusion flow to some degree.
We group \tcode~records by the number of \textit{close neighbors} of the receivers, 
and compare their average CI across different product categories,
as illustrated in \figref{fig:ob:all}(c).
We find that, in general, the more close neighbors the receiver has, the more likely it is that the receiver will buy the item;
i.e., the average CI increases from 18.8 to 25.3 as the $x$-$axis$~ becomes large.
We can therefore conclude that users would have a higher CI on an item if many of their neighbors are in the diffusion flow of this product,
 (i.e., receive the \tcode~message of this item and buy it).
Our observations are consistent with ~\cite{portrait}, who found that users with social connections may have similar shopping behaviors.
In conclusion, \textbf{the network structure of interest diffusion networks may help us in modeling user purchasing behaviors.}

\subsection{Temporal dynamics of \tcode}
\label{subsec:observe:temporal}
Finally, we tackle the temporal dynamics of \tcode~diffusion:
\textbf{Does the senders' past receiving history do have impact on receivers' purchasing? (Q3)}



In the literature, a diffusion cascade can be defined as a information flow, as item $p$ flows by $u_1$->$u_2$->$u_3$->...->$u_k$. 
To simplify the diffusion process, we only look at the 2-hop cascades that end with the user $u_3$ in the above example.
For convenience of explanation, we refer to the target user as $u_3$, and his/her 1-hop and 2-hop neighbor on an in-edge of a diffusion path as $u_2$ and $u_1$ respectively.
We aim to figure out 1) Does the 1-hop neighbor($u_2$)'s receiving records have impact on the receiver($u_3$)'s purchasing and 2) How does the impact changes over time? 

We present the observational result in \figref{fig:ob:all}(d), where x-axis $\delta$ donate the time gap(weeks) between the time when $u_2$ received the same \tcode~ and the current \tcode~ sent to $u_3$, note that if $u_2$ does not receive the same \tcode~ before, the $\delta$ is set to 0. while y-axis donate the \ratios.
The first of our finding is that the average CI of \tcode~ with $\delta=0$ (no 2-hop neighbors) is 19.1$\pm$4.32, which is lower than the case when 2-hop neighbors exists($\delta$>0).
Moreover, there are no significant differences in the averaged CI between different time gaps $\delta$:
the results of remaining groups ($\delta=2,3,4$ weeks) are in 19.5$\pm$4.49, 19.2$\pm$4.59, and 19.99$\pm$4.16 respectively. 
However when it comes to the case in which the sender buys the item later (the blue bars),
we can observe that the average CI is much higher than that of the red bars,  
even if the previous \tcode~sharing occurs more than four weeks ago (20.4$\pm$4.07, vs. 19.99).
This suggests that \textbf{both the long- and short-term temporal dynamics of \tcode~diffusion do indeed exist.}
This is consistent with the common-sense conclusion that (1) the more frequently \tcode~ messages are spread within a short time span, 
the more likely it is that the item has great popularity with a higher CI;
(2) alternatively, a user $u$ will probably share some product with others long after she receives, and after spending some time making sure that the item deserves the recommendation.
Thus, the temporal dynamics of \tcode~diffusion could be an important factor in influencing user purchasing behaviors.

\section{Model: \textit{\modelname~}}
\label{sec:model}

\begin{figure*}[ht]
	\centering
	\includegraphics[width=1.0\textwidth]{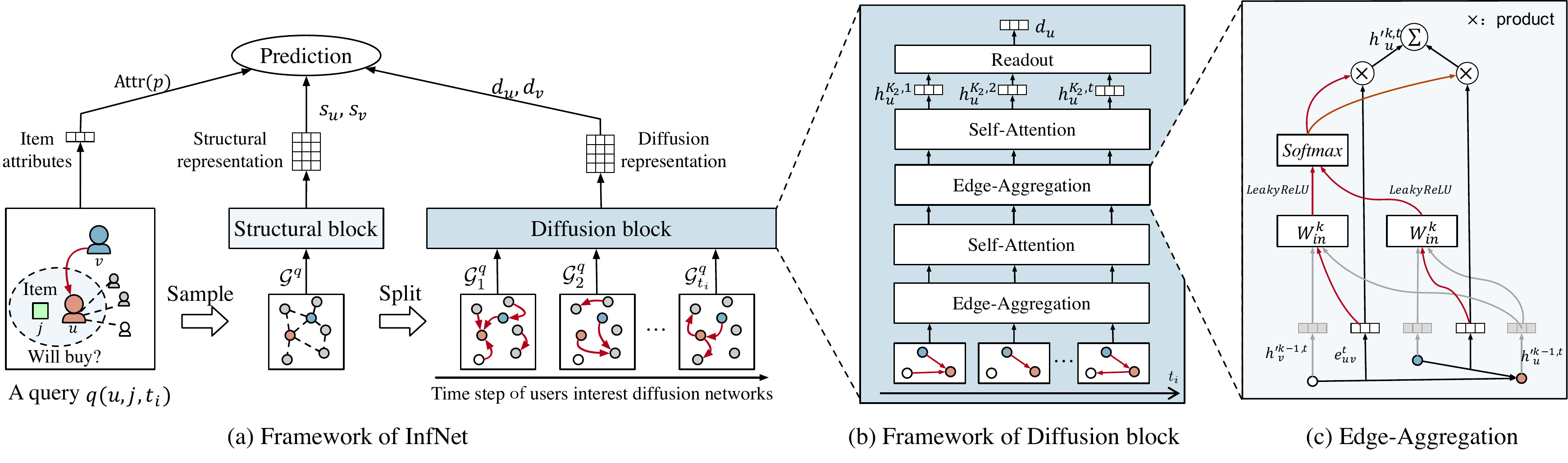}
	\caption{Framework of \modelname.
		\small (a) is the overview of \modelname, (b) is the framework of diffusion encoder in \modelname. (c) is the detail of edge-level aggregation. Note, in (c), user's hidden states only used to count attention weight. 
		}
	\label{fig:model:overview}
\end{figure*}

In this section, we introduce the proposed model \modelname~ for modeling the structure and temporal dynamics of \tcode~ diffusion in the interest diffusion networks. 
Given the input of a purchasing query at a specific time step that contains the target user and item,
our framework first builds a sequence of ``dynamic sub-networks'' for each query,
after which we apply two different attention-based aggregation strategies on dynamic node and edge attributes to learn users' hidden representations. 
Finally, the user's embedding, along with the item features are fed into the outer classifier,
to predict whether the target user will purchase the item during the given time span. 
An overview of the framework is presented in \figref{fig:model:overview}.


\subsection{Preprocessing: Sub-graph construction}
\label{subsec:model:sub-graph}
For each purchasing query $q=(u, p, t_i)$ and the entire dynamic interest diffusion network $\mathcal{G} = \{\mathcal{G}_1, \cdots, \mathcal{G}_{t_i}\}$,
we build a sequence of sub-graphs $\mathcal{G}^q$ of $\mathcal{G}$ to conduct the end-to-end graph classification, i.e., the purchasing prediction of query $q$.
The process of constructing the sub-networks can be described as follows :
\begin{itemize}[leftmargin=*]
	\item \vpara{Seed node sampling:} We define the target user $u$ and its 1-hop neighbors $\mathcal{S}_u(p)$ as the set of seed nodes,
	denoted as $\{u\} \bigcup \mathcal{S}_u(p)$, where $\mathcal{S}_u(p)$ is the set of \textit{sender nodes} that shares item $p$ with $u$.
	\item \vpara{Sub-network construction:} Given the set of seed nodes, we adopt a breadth-first search (BFS) strategy to sample the nodes
	from each original diffusion network $\mathcal{G}_t (1\le t \le t_i)$, by searching for two-hop neighbors, i.e., the search deepth is set to two;
	and moreover, an edge exists between two nodes $(u, v)$ in $\mathcal{G}_t^q$ if the user $u$ shares some product $p$ with $v$ at time step $t$.
	\item \vpara{Node and edge attributes:}
    The node features of a user $u$, denoted as $\mbox{Attr}(u)$, consists of two parts: 
    the first part is the user's purchasing histories which are independent from the diffusion networks, 
    i.e., the price she pays for different products in the past year;
    the other part is a one-hot vector, indicating whether this node is the target user in the query or it is among the seed nodes,
    which can help the model better identify the nodes that it should focus on. 
    As for the edge features,
    we first define the item feature $\mbox{Attr}(p)$ as a one-hot vector, where each bin represents a specific range of prices.
    Then for an edge $(u,v,t)$, its features $e_{uv}^t$ is summed over features of all items that are sent from the user $u$ to $v$ during the period $t$. 
    Note that edge features may change across queries at different time, since for a specific time span $t$, 
    we only aggregate features of items that are shared during $t$.
\end{itemize}

\subsection{\modelN}
As \figref{fig:ob:all} (b) shows, the local graph structure in the diffusion network is an important factor to infer whether \tcode~sharing can prompt purchases.
Therefore, we use a \modelN~to learn the users' embeddings to capture such structural characteristics. 
\modelN~can be implemented by any GNNs with node aggregation strategy, which can embed nodes' structural information~\cite{gin}. It can be denoted by a uniform Message-Passing framework:
for a specific user $u$, the first layer is a non-linear transformation on user attributes, and output the initial hidden representation $x^u$:
%
\begin{equation}
x^k_u = \mbox{Combine}(x^{k-1}_u, \mbox{Aggr}({x^{k-1}_v}|v \in \mathcal{N}_u))
\label{eq:messagepassing}
\end{equation}
where $k$ is the layer index,
$\mbox{Combine}(\cdot,\cdot)$ combines hidden states of $u$ with her neighbors, 
and $\mbox{Aggr}(*)$ denotes the aggregation function (sum or mean operation). 
By this step, the representation vector $x_u^k$ is expected to contain the information of both $u$ and her neighbors. 
After stacking $K_1$ layers of \eqnref{eq:messagepassing},
the output is the final structural embedding of the user $u$, 
denoted as $s_u = x^{K_1}_u$.

\subsection{\modelE}
While \modelN~ only aggregates on the node attributes, edge features encode much more information of \tcode~diffusion:
as \figref{fig:ob:all} (c) and (d) suggest, 
if we want to better predict the user $u$'s purchasing decision on the item $p$, 
we may need to know how many close neighbors of $u$,
as well as whether there exist some users who have shared $p$ to $u$ having purchased this item.
Moreover, the temporal information of \tcode~ diffusion can also help us capture both short- and long-term influence of such social interactions on users' purchasing behavior.
However, existing methods rarely consider the dynamic item diffusion among users: most of them apply frameworks similar to \eqnref{eq:messagepassing},  which is unsuitable for a \tcode~ diffusion network. 
Consider a situation, in which user $v$ has received many \tcodes~of computers and clothes, and sends a clothing \tcode~to $u$. 
The diffusion from $v$ to $u$ pertains only to clothes; 
therefore, node information of $v$ related to computers may evidently introduces noise that hampers the prediction of $u$'s purchase behavior. 
So we further design \modelE~ to fully model the \tcode~diffusion via edge-level aggregations and learn users' final representations.

Inspired by real-world situations and our observations (\secref{sec:observe}), 
we design a special multi-layer aggregation strategy. 
The key idea is that, rather than directly aggregating neighbors' node features, 
we first learn the weight of each edge, then aggregate the edge features between the user and her neighbors to get an updated representation. 
Through the attention mechanism, neighbors' node representations will affect the nodes' final representation in an implicit way, and it can avoid noisy aggregations. 


Formally, let $e_{uv}^t$ denotes the edge feature of user $v$ sent to user $u$ at time $t$. 
To capture the dynamics of \tcode, 
we independently aggregate the attributes of edges connected with the user to get the hidden state of each time step, denoted as ${h'_{u}}^{k,t}$, where $k$ is the layer number. 
Each hidden state ${h'_{u}}^{k,t}$ only contains all interplay on social networks involving $u$. 
We then utilize a sequence encoder to combine users' hidden states in different time steps to get the dynamic representations of users. 
In detail, the same user at different time step has the same initial hidden state $h^{0,t}_u$, which is obtained by a non-linear transformation of user attributes $\mbox{Attr}(u)$;
then for user $u$ in $k^{th}$ layer at time step $t$, we conduct separate aggregations of users' in- and out-edges and obtain two representations $In_u^{k,t}$ and $Out_u^{k,t}$. They represent the information users receive and send. We combine them to get the hidden state $h'^{k,t}_u$ at time step $t$:
\begin{equation}
In^{k,t}_u = \sum_{v \in \mathcal{S}_u^t} {\mbox{ATT}^k_{in}(h^{k-1,t}_u,h^{k-1,t}_v,e_{uv}^t)e_{vu}^{t}}
\label{eq:attention-in}
\end{equation}

\begin{equation}
Out^{k,t}_u = \sum_{v \in \mathcal{R}_u^t} {\mbox{ATT}^k_{out}(h^{k-1,t}_u,h^{k-1,t}_v,e_{vu}^t)e_{vu}^{t}}
\label{eq:attention-out}
\end{equation}

\begin{equation}
h'^{k,t}_u = [In^{k,t}_u,Out^{k,t}_u]
\end{equation}
Here $\mbox{ATT}^*_*(*)$ is an attention function that takes the user's hidden states in the last layer, as well as edge attributes as inputs, and output the attention weights of the edge attributes. 
It fuses these three inputs, then employs a non-linear transformation to obtain the scalar values. 
The attention mechanism can be clarified as:
\begin{equation}
w_{uv}^{k,t} = \mbox{LeakyReLU}(W_{in}^k[h_u^{k-1,t},h_v^{k-1,t},e_{uv}^t])
\end{equation}
\begin{equation}
\mbox{ATT}^k_{in}(h^{k-1,t}_u,h^{k-1,t}_v,e_{uv}^t) = \frac{exp(w_{uv}^{k,t})} {\sum_{c\in\mathcal{S}_u^t}{exp(w_{uc}^{k,t})}}
\label{func:softmax}
\end{equation}
where $W_{*}^k \in \mathbb{R}^{1\times 3c}$, and we adopt a softmax function ~\cite{gat} on edges to make the summed weights of all $u$'s neighbors equal to 1. 
This operation can alleviate over-fitting and facilitate fair comparison of various edges. 
The out-edge attention is the same as that of in-edge, except that in \eqnref{func:softmax}, it is summed over $\mathcal{R}_u^t$ rather than $\mathcal{S}_u^t$.

After obtaining the hidden state at each time step $t$, i.e.,
a sequence of representations of different time steps for each user $u$,
we use a sequence encoder to capture the temporal dynamics:
\begin{equation}
h^{k,t}_u = \mbox{Enc}(h'^{k,i}_u | i<=t)
\label{eq:seqence encoder}
\end{equation}
The $\mbox{Enc}(*)$ can be a simple operator such as Mean-pooling or a more complex operator such as GRU~\cite{gru}, Masked self-attention~\cite{vaswani2017attention}, etc. 
However, as this aspect is not our main focus, here we do not expand on it in detail, and will analyze the performance of different encoding methods in \secref{sec:exp}. 

After $K_2$ layers of \modelE, we obtain the final dynamic representation $d_u = \mbox{Readout}(h^{K_2,t}_u|_{t=1}^{t=t_i})$, where $\mbox{Readout}(*)$ denotes readout the overall representation of sequence. For example, we can directly readout the representation in last time step in GRU~\cite{gru}, or readout by sum with weights in Self-Attention ~\cite{dosovitskiy2020image}.

\subsection{Model learning}
After we obtain the users' structural and \tcode~ diffusion representations, 
we combine them to predict the users' purchasing behavior: 
whether user $u$ will purchase item $p$, namely query $q=(u,p)$.
We focus not only on $u$'s representation and item $p$'s attributes, but also consider the users who send item $p$ to $u$:
\begin{equation}
g_q = \sum_{v \in \mathcal{S}_u(p)}{[s_v,d_v]}
\label{func:final}
\end{equation}
Here $\mathcal{S}_u(p)$ is the set of seed users (\secref{subsec:model:sub-graph}), in which all users are $u$'s neighbors who send item $p$ to $u$.
We then combine these three parts and make the predictions through a simple output layer:
\begin{equation}
y' = \sigma([d_u,s_u,g_q, \mbox{Attr}(p)])
\end{equation}
where $\mbox{Attr}(p)$ denotes the features of item $p$.

In general, the downstream task can be regarded as a binary classification problem, thus we use \textit{cross-entropy} as the loss function:
\begin{equation}
\mathcal{L}(\Theta) = \sum_{q_i \in Q} -y_i \mbox{log}y_i'+(1-y_i)\mbox{log}(1-y_i')
\end{equation}
Here, $\Theta$ denotes all learnable parameters in \modelname, $Q$ is the set of all queries, and $y_i$ is the ground truth of the query $q_i$.
\section{Experiments}
\label{sec:exp}
\subsection{Experimental setup}
\vpara{Prediction Datasets.} 
We select \mbox{\tcode~}sharing messages from six categories for our experiments.
In more detail, in order to formulate the purchasing prediction problem with clear input and avoid data selection biases, we collect 30,000 \mbox{\tcode~}records for each category on each day
during the three days following the observational time window in \mbox{\secref{sec:observe}}, with a total of 540,000 samples,
referred to as ``\textit{the prediction datasets}''.

We regard the receiver and the item of each \mbox{\tcode~}record as a purchasing query $q$, which asks whether the user will buy this item within a specific time span.
We then inquire into the purchasing logs during the time period to label these queries (i.e., whether or not the user will buy the item).
For each query sample, the input for our model is the dynamic diffusion sub-network defined in \secref{subsec:model:sub-graph},
within the given time period.
More specifically, we divide the whole time span into four equal segments, meaning that there are a total of four diffusion sub-graphs constructed for each query with one week as a time step.
To keep the offline experiment consistent with the online situation, where we could not obtain any information ahead of the querying time when making predictions,
we randomly split the datasets on the first two days into training and validation sets
with a ratio of 7:3,
with all queries on the third day are used for testing.
Overall statistics of the datasets for the item-level purchasing prediction task are presented in \tableref{tbl:exp:data}.
\begin{table}[t!]
	
	\caption{Statistics of the prediction datasets.
	\small
	Due to data security requirement, we rename the categories into ranked ids.
	\normalsize
	}
	\label{tbl:exp:data}
		\begin{tabular}{c|c|c|c|c}
			\hline
			Category & \# Users & \# Items & \# Edge &\# Purchase \\ \hline 
			Cate\#1   & 1,261,486      & 33,184   & 2,730,610   & 21,633 \\ \hline 
			Cate\#2  & 1,289,775      & 38,776   & 2,832,443   & 29,538  \\ \hline 
			Cate\#3  & 664,732      & 5,196   & 1,396,300   & 80,278 \\ \hline 
			Cate\#4   & 997,957      & 19,632   & 2,160,293   & 11,657 \\ \hline 
			Cate\#5  & 891,134      & 74,487  & 1,911,424  & 9,181  \\ \hline 
			Cate\#6  & 1,267,316      & 47,920   & 2,806,235   & 41,008 \\ \hline 
			All  & 6,372,400      & 21,9195   & 13,837,305   & 122,739 \\ \hline 
		\end{tabular}
	
\end{table}

\begin{table*}[t!]
	
	\caption{Experimental results of item-level purchasing prediction tasks.
		\small
		We use $^*$ to denote the best result among all models,
		while an underline indicates the best performance among the baseline methods.
	}
	\label{table:performance}
		\begin{tabular}{c|p{0.75cm}p{0.75cm}|p{0.75cm}p{0.75cm}|p{0.75cm}p{0.75cm}|p{0.75cm}p{0.75cm}|p{0.75cm}p{0.75cm}|p{0.75cm}p{0.75cm}}
			\toprule
			\multirow{2}{*}{\textbf{\diagbox{Methods}{Datasets}}}  & \multicolumn{2}{c|}{\textit{Cate\#1}}        & \multicolumn{2}{c|}{\textit{Cate\#2}}          & \multicolumn{2}{c|}{\textit{Cate\#3}}        & \multicolumn{2}{c|}{\textit{Cate\#4}} & \multicolumn{2}{c|}{\textit{Cate\#5}}  & \multicolumn{2}{c}{\textit{Cate\#6}}   \\
			 & \textit{ROC} & \textit{PR} & \textit{ROC} & \textit{PR} & \textit{ROC} & \textit{PR} & \textit{ROC} & \textit{PR} & \textit{ROC} & \textit{PR} & \textit{ROC} & \textit{PR}\\ \midrule

        LR &0.607&0.310&0.565&0.371&0.523&0.048&0.641&0.200&0.636&0.152&0.567&0.479\\
        BPR &0.599&0.317&0.550&0.367&0.598&0.057&0.614&0.164&0.562&0.134&0.549&0.492\\\midrule
    
        SR-GNN &0.600&0.318&0.569&0.382&0.599&0.059&0.603&0.187&0.558&0.126&0.558&0.495\\
        MGNN-SPred & 0.604&0.319&0.554&0.370&0.573&0.054&0.637&0.190&0.559&0.126&0.567&0.514\\
        GCE-GNN &0.615&0.330&0.573&0.389&\underline{0.613}&0.064&0.655&0.206&0.561&0.127&0.569&0.515
\\\midrule
        
        EATNN  &0.508&0.246&0.512&0.340&0.508&0.042&0.507&0.120&0.512&0.107&0.505&0.450\\
        GraphRec &0.547&0.277&0.562&0.375&0.542&0.048&0.623&0.171&0.573&0.132&0.515&0.474\\
        DiffNet &\underline{0.639}&\underline{0.357}&\underline{0.582}&\underline{0.392}&0.612&\underline{0.071}&\underline{0.687}&\underline{0.233}&\underline{0.654}&\underline{0.187}&\underline{0.612}&\underline{0.598}\\ \midrule
        

        \modelname-S &0.650&0.368&0.620&0.428&0.587&0.065&0.669&0.240&0.704&0.244&0.635&0.578\\
        \modelname &\textbf{0.671*}&\textbf{0.409*}&\textbf{0.644*}&\textbf{0.466*}&\textbf{0.643*}&\textbf{0.100*}&\textbf{0.726*}&\textbf{0.319*}&\textbf{0.742*}&\textbf{0.323*}&\textbf{0.657*}&\textbf{0.606*}\\
        \bottomrule
		\end{tabular}
\end{table*}

\vpara{Baselines.} We compare our proposed model, \modelname, with several groups of state-of-the-art baselines:

\begin{itemize}[leftmargin=*]
\item{Traditional methods.} 
We first compare our model with a classical machine learning method: Logistic Regression (LR) based on feature engineering.
We also choose one of the most popular frameworks in the item predictions, Bayesian Personalized Ranking (BPR)~\cite{rendle2012bpr},
which proposes a maximum posterior estimator as the generic optimization criterion for personalized ranking.

\item{Session-based methods.} 
Another line of online purchasing predictions works are the session-based models.
Since \tcode~message sharing sequence can be naturally regarded as user actions within a session,
we choose three popular session-based baselines: SR-GNN~\cite{wu2019session}, MGNN-SPred~\cite{wang2020beyond} and GCE-GNN~\cite{wang2020global}.

\item{Social recommendation methods.} 
To verify the effectiveness of \modelname~ in capturing social dynamics across the interest diffusion networks, 
we compare three popular social recommendation frameworks: EATNN~\cite{chen2019efficient},  GraphRec~\cite{fan2019graph} and DiffNet~\cite{wu2020diffnet++}. 
EATNN introduces attention mechanisms to model users' preferences and assigning a personalized transfer scheme for each user.
GraphRec incorporates user-item and user-user graphs, 
and uses attentions to model the importance of different social relationships.
DiffNet considers the social diffusion by applying GCN~\cite{gcn} on the social networks and SVD++ to the item recommendations.

\end{itemize}

\vpara{Evaluation metrics and implementation details.}
We evaluate the predictive performance of \modelname~ and baseline methods in terms of AUC for the Precision-Recall (PR) and ROC curves, as these metrics are widely applied in recommendation systems~\cite{davis2006relationship}.

In order to fairly compare the performance of different models, 
and to adjust to our problem settings,
we make some modifications to the deployment of baselines: 
\begin{itemize}[leftmargin=*]
	\item{LR based on feature engineering.}
	We carefully extract three categories of features:
	(1) cost-related features, which contain the price of the product in the query and the total spending of the user in the past 30 days; 
	(2) graph-related features, which contains the in- and out-degree of users in the diffusion network; 
	(3) features of users' historical \tcode~sharing records, i.e., the information about products the user has shared and received. 
	
	\item{Session-based methods}. 
	We regard users’ \tcode~sharing records during the given time span as a session (these are used to extract node and edge features in \modelname),
	then add the price-related features when encoding the item information as a one-hot vector. 
	
	\item{Social-based recommendations}. 
	For all GNN-based models, the input is the static interest diffusion network by aggregating all dynamic sub-networks in \modelname, which are constructed from users' previous sharing records.
	Moreover, since above-mentioned methods are based on static graphs and features,
	to fairly compare their performance with \modelname, we design the \modelname-S, where all user and item features are completely consistent with the social-based recommendation baselines.
\end{itemize}

\begin{table*}[ht!]
    \caption{In-deep analysis on \modelname.
 	    \small
    	``-'' means we remove the corresponding part in \modelname. The reported metric is AUC\_{PR}.}
 	\normalsize
    \begin{tabular}{c|c||c|c|c|c|c|c}
    \toprule
    \textbf{Analytical Category}                             & \textbf{Variant}          & Cate\#1 & Cate\#2 & Cate\#3 & Cate\#4 & Cate\#5 & Cate\#6      \\\midrule
                                                    & \textbf{None}              & 0.393&0.460&0.091&0.312&0.292&0.597  \\
    \multirow{2}{*}{(a) Sequence Encoder Selection} & \textbf{Mean-Pooling}              & 0.402&0.464&0.086&0.308&0.313&0.603  \\
                                                    & \textbf{GRU}              & 0.403&0.466&0.094&0.316&0.322&0.606  \\\midrule
                                                    & \textbf{-User feature}    &  0.407&0.464&0.086&0.297&0.305&0.599 \\
    (b) Effect of \tcode.                             & \textbf{-Item feature}    & 0.399&0.456&0.092&0.289&0.290&0.578  \\
    \multicolumn{1}{l|}{}                            & \textbf{-Taocode feature} & 0.355&0.406&0.079&0.282&0.236&0.577 \\\midrule
    (c) Attention mechanism.                        & \textbf{-Attention}       & 0.399&0.456&0.092&0.289&0.290&0.578   \\\midrule
    (d) Structural encoder                      & \textbf{-\modelN}     & 0.403&0.464&0.085&0.313&0.318&0.587
  \\\midrule 
    \multicolumn{2}{c||}{\textbf{\modelname}}                                      & \textbf{0.409}&\textbf{0.466}&\textbf{0.100}&\textbf{0.319}&\textbf{0.323}&\textbf{0.605} 
	\\ \bottomrule
    \end{tabular}
\label{tbl:ablation}
\end{table*} 
Besides, we use the Adam~\cite{kingma2014adam} optimizer with a learning rate of 0.001, and the batch size is set as 512. 
In \modelname, we search the hidden sizes $c$ in the range [16, 32, 64, 128], and $c=64$ reaches the best performance.
For the GNNs in \eqnref{eq:messagepassing}, 
we have tried GIN \cite{gin}, GCN \cite{gcn} and GAT \cite{gat}. 
We find that there are little differences on the performance of different GNNs models, and choose GAT whose average performance is the best among all candidates.
For the sequence encoder in \eqnref{eq:seqence encoder}, we select Masked Self-Attention (\secref{subsec:exp:ablation}), where the Readout function is similar to \cite{gat}.

\subsection{Experimental results}
We present the performance of all methods on item-level purchasing prediction task on the \tcode~ dataset in \tableref{table:performance}. 
Overall, \modelname~ achieves the best performances on all six categories:
\modelname~ is expected to be able to capture the spread of product interests along with both structural and temporal dynamics of \tcode~diffusions, 
and brings in an averaged relative performance increase of 4.6\% and 6.3\% in terms of AUC\_{PR} and AUC\_{ROC} respectively compared with the best baseline method, 
demonstrating the effectiveness of our approach for purchasing predictions:

\begin{itemize}[leftmargin=*]
    \item In detail, although BPR only utilizes the information of user-item interactions while LR infers purchasing behaviors using only the user and item features,
their performances are competitive with many baselines, which indicates that handcrafted features are crucial in item-level purchasing predictions.

    \item Most of the session-based methods focus primarily on the correlations between items across different time steps. 
    However, due to the huge number of possible action sequences (\tcode~sharing) within a session,  
    this group of methods could be easily over-fitted on the training set, and they also ignore some key properties of users' purchasing behavior. 
    And when the scale of the dataset is relatively large (Cate\#1, \#2, \#5 and \#6), session-based models achieve only a limited improvement over LR and BPR. 

    \item Social recommendation methods are the most appropriate to our experimental setting, and most similar to \modelname, as they not only consider the user-item interactions but also model the social dynamics among users. 
    For instance, EATNN uses a whole data-based optimization strategy for neural models, 
    GraphRec constructs both item-item and user-user graphs based on users' social interactions,
    and DiffNet uses a fuse layer to combine users' and items' features, 
    then apply GCN to model the interest diffusion in the social network.
    However, EATNN and GraphRec only consider the structural properties of \tcode~ diffusion networks while ignore the attributes on nodes and edges,
    which may lose too much information of \tcode~messages.
    Furthermore, EATNN aims to rank all items for each user without negative sampling, bringing in performance drops on purchasing predictions of each single item.
    Although DiffNet achieves the best performance among all baselines, 
    it does not consider the different influence of a user on her different neighbors, as well as the various impacts of \tcode~records with different items.
    Therefore, those social recommendation methods cannot make accurate predictions on item-level purchases based on \tcode~ diffusion.
\end{itemize}

\subsection{In-depth analysis of \modelname}
\label{subsec:exp:ablation}
\modelname~ consists of a structural encoder based on a node-aware GNN,
and a dynamic \tcode~ diffusion encoder with attention-based edge aggregations and a sequence encoder. 
To investigate whether these encoders or attention mechanisms actually work and how they influence the performance of \modelname,
we conduct ablation studies by removing each part of key component,
and evaluate the performances of these variants of \modelname~ in \tableref{tbl:ablation}:

\begin{itemize}[leftmargin=*]
     \item \textbf{Sequence Encoder Selection}. We adopt Masked Self-attention as the sequence encoder in \eqnref{eq:seqence encoder},
     where elements in each time step are encoded only with reference to the previous time steps. 
     Besides, there are many other deep architectures designed to encode sequences, such as RNN-based methods. 
     In order to verify the influence of self-attention, 
     we compare with the model replacing with GRU. 
     In addition, to validate the necessity of sequence encoding, we further compare our methods with simple Mean-Pooling or even None (use only the output of last time step without any encoder).
     According to the results shown in \tableref{tbl:ablation}(a),
     it is important to capture the temporal dynamics at different time steps by a sequence encoder,
     and simply using mean-pooling is unable to capture long- and short-term dependence, thus it cannot achieve a good performance on some time-sensitive categories (Cate\#3 and \#4). 
     We choose Self-Attention because it is better at capturing long-term dependencies and achieves superior performance than GRU. 
 
    \item \textbf{Effect of \tcode}. In order to intuitively demonstrate the importance of \tcode~, we remove different part of input features. As shown in \tableref{tbl:ablation}(b), 
    masking \tcode~ features results in the sharpest drop in performance. This indicates that \tcode~ is a key factor to predict users' purchasing behavior, so \modelname~can model \tcode~diffusion better for purchase prediction.
    
    \item \textbf{Attention mechanism}. 
    In \eqnref{eq:attention-in} and  \eqnref{eq:attention-out}, \modelname~ assigns dynamic weights to each edge via graph attentions. In order to clarify the importance of the attention mechanism, 
    we draw a comparison between \modelname~ and its variant without attention. 
    As \tableref{tbl:ablation}(c) shows, the performances of \modelname-Attention are worse on all six categories. with a maximum drop of 10.2\% on Cate \#3. Therefore, we conclude that the attention mechanism is essential for modeling \tcode~ interest diffusion. 

    \item \textbf{Structural encoder}. \modelN~ is designed to encode user's structural information. To evaluate whether this module can help us infer users' purchase behavior, we compare with \modelname~ without this component. 
    Again, as in \tableref{tbl:ablation}(d), 
    removing \modelN~ leads to a 2.3\% decline in performance.
    \end{itemize}
    To summarize, according to the above discussions: (1) The sequence encoder is a necessary component, and self-attention is the best choice for \modelname. (2) \tcode~ is the most informative among the three input features of \modelname. (3) Attention mechanism can help \modelname~ better model \tcode~interest diffusion. (4) Structural information also benefits in improving the proposed model.


\begin{table}[ht]
\caption{Comparison on cold-start problem. C\# is short for Cate\# and using bold to highlight the better improvement.}
 	    \small
 	\normalsize
\begin{tabular}{c|c||c|c|c|c|c}
    \toprule
Group                  &   Method &\textit{C\#1}&\textit{C\#2}&\textit{C\#3}&\textit{C\#4}&\textit{C\#5}\\ \midrule
\multirow{3}{*}{\textit{Cold.}} 
& DiffNet &0.331&0.332&0.061&0.159&0.143 \\\cline{2-7}
& Ours&0.392 &0.381&0.113&0.230&  0.276\\\cline{2-7}
& \textbf{Impv.}& \textbf{18.4\%}& 14.8\% &\textbf{85.2\%}&\textbf{44.7\%}&\textbf{93.0\%}
\\\hline\hline
\multirow{3}{*}{\textit{Warm.}} 
& DiffNet & 0.358&0.396&0.074&0.244&0.188   \\\cline{2-7}
 & Ours&0.410&0.472 &0.086&0.329&  0.329   \\\cline{2-7}
& \textbf{Impv.}&14.5\%&\textbf{19.2\%}&16.2\%&34.8\%& 75.0\%  \\
                         \bottomrule
\end{tabular}
\label{exp:case}
\end{table}
\subsection{Cold-start challenge}

   As we observed in \secref{subsec:exp:ablation}, \tcode~ features are important for purchase predictions. 
   To further verify the power of \modelname, we conduct an analysis on the challenge which traditional recommendation system suffered, namely the \textbf{cold-start problem}~\cite{schein2002methods}. There are two distinct categories of cold start: product cold-start and user cold-start, while here we focus on user cold-start problem.
   
   In our setting, we denote \textit{Cold} for the group of users who have no purchase records in the past and \textit{Warm} on the contrary.
   We then compare \modelname~ with DiffNet (the best baseline), shown in \tableref{exp:case}.
   Although both models suffer in cold-start dilemma (perform worse under the \textit{Cold} setting compared with \textit{Warm}), 
   \modelname~ shows a better improvement than DiffNet with cold-start.
   More specifically, except for Cate\#2, \modelname~'s improvements over DiffNet on \textit{Cold} are significantly higher than that of \textit{Warm},
    It shows that \modelname~ is a potentially better approach to tackle the cold-start problem.

\section{Related work}
\vpara{Purchase prediction in the recommendation system.}
Online shopping has been investigated since the early stages of the Web. Purchase behavior prediction in e-commerce is a classic but difficult problem. 
Many previous works studies online shopping behaviors based on features of users and products~\cite{sismeiro2004modeling,portrait,viral}, such as item price and user credit records, etc.
Besides, traditional collaborative filtering~\cite{sarwar2001item} and deep learning techniques are used to predict the probabilities of which item will be purchased by the user. 
Some of them try to extract the implicit correlations between users and items~\cite{koren2008factorization,he2017neural,wang2019neural,he2020lightgcn}.
Another kind of methods focus on the click and purchase records of users, which treat the click/purchase records as a sequence and encoding them with a sequence encoder such as RNNs~\cite{hidasi2015session}, and GNNs~\cite{wu2019session,wang2020beyond,wang2020global}. 
The goal of these methods is digging out the regular pattern of item purchases. 

\vpara{Social recommendations.} 
Due to the potential value of social relations and user interactions in the recommendation systems, social recommendation has attracted increasing attention~\cite{tang2013social}. As theories from the domain of social science show the homophily and influence~\cite{bakshy2012role} between social relations, various approaches are proposed to build social recommendation systems such as trust ensemble~\cite{ma2009learning}, trust SVD~\cite{guo2015trustsvd}, and social regularization~\cite{ma2011recommender}. 
More recently, deep learning methods as well as transfer learning~\cite{chen2019efficient} and GNNs~\cite{fan2019graph,wu2019neural,wu2020diffnet++,xu2019relation} are also adopted to solve the problem.

\vpara{Deep learning on graphs.}
Recently, GNNs have achieved state-of-the-art performance on many tasks~\cite{gcn,gat,gin,wang2021modeling}. 
Based on the Message-Passing framework, many works have expanded the capabilities to tackle different types of graph, 
such as heterogeneous graphs~\cite{hgt} and dynamic graphs~\cite{hu2021time,zhou2018dynamic}. These improvements make GNNs better compatible for purchase prediction in real-world applications~\cite{fan2019metapath,song2019session}. 
\section{Conclusion}
\label{conclude}
This paper investigates the influence of information diffusion on user purchasing behaviors in an online shopping platform. 
Taking \tcode~
as a case study, 
we collect a large-scale real-world dataset that includes over 100M \tcode~sharing records. 
Based on the dataset, 
we conduct empirical observations to explore the intrinsic properties of \textit{product interest diffusion} on \taobao, 
finding that \tcode~ diffusion exerts a strong influence on user purchasing behaviors,
while both structural and temporal dynamics of diffusion networks play a key role in such correlations. 
Inspired by these observational insights, we design an end-to-end GNN-based framework, dubbed \modelname,
to model the product interest diffusion via \tcode.
More specifically, we apply both graph- and sequence-level attention mechanisms to capture the dynamics of user interests on different products at different time steps.
On the item-level purchasing prediction task on the real-world \tcode~diffusion datasets from six different product categories, 
\modelname~achieves significantly better performance compared with state-of-the-art baselines;
moreover, ablation studies demonstrate the additional predictive power introduced by the careful design of our structural and temporal components.
We hope this work will bring insights for user purchase behavior modeling,
especially from the perspective of user interest diffusion and social interactions.

\vpara{Acknowledgments.} 
This work is supported by the National Key Research and Development Project of China (No. 2018AAA0101900) and a research funding from Alibaba Group. 
Yang Yang's work is supported by Tongdun Technology. 

\bibliographystyle{ACM-Reference-Format}
\balance
\bibliography{InfoNet}

\end{document}